\documentclass[9pt]{article}
\usepackage{spconf,amsmath,graphicx}
\usepackage{url}
\usepackage{colortbl}
\usepackage{array}
\usepackage{booktabs}
\usepackage{multirow}
\usepackage{makecell}
\usepackage{hyperref}
\usepackage{xcolor}
\usepackage[tableposition=top]{caption}
\makeatletter
\gdef\@email{}
\makeatother
\usepackage{cite}
\usepackage[bottom, splitrule]{footmisc}

\title{DFADD: The Diffusion and Flow-Matching based Audio Deepfake Dataset}
\name{\parbox{\textwidth}{\centering
\begin{tabular}{c}
Jiawei Du$^{1*}$, I-Ming Lin$^{1*}$, I-Hsiang Chiu$^{3*}$, 
Xuanjun Chen$^{2}$, Haibin Wu$^{2}$, Wenze Ren$^{2}$, \\
Yu Tsao$^{4}$, Hung-yi Lee$^{2}$, Jyh-Shing Roger Jang$^{1}$\thanks{$^*$equal first contribution}
\end{tabular}}}

\address{$^1$Department of Computer Science Information Engineering, National Taiwan University \\
$^2$Graduate Institute of Communication Engineering, National Taiwan University \\
$^3$Department of Electrical Engineering, National Taiwan University \\
$^4$Academia Sinica, Taiwan}

\begin{document}
%
\maketitle

\begin{abstract}
Mainstream zero-shot TTS production systems like Voicebox and Seed-TTS achieve human parity speech by leveraging Flow-matching and Diffusion models, respectively.
Unfortunately, human-level audio synthesis leads to identity misuse and information security issues. 
Currently, many anti-spoofing models have been developed against deepfake audio. 
However, the efficacy of current state-of-the-art anti-spoofing models in countering audio synthesized by diffusion and flow-matching based TTS systems remains unknown. 
In this paper, we proposed the Diffusion and Flow-matching based Audio Deepfake (DFADD) dataset. 
The DFADD dataset collected the deepfake audio based on advanced diffusion and flow-matching TTS models. 
Additionally, we reveal that current anti-spoofing models lack sufficient robustness against highly human-like audio generated by diffusion and flow-matching TTS systems. 
The proposed DFADD dataset addresses this gap and provides a valuable resource for developing more resilient anti-spoofing models.
\end{abstract}

\begin{keywords}
dataset, deepfake detection, anti-spoofing, text-to-speech
\end{keywords}

\section{Introduction}
Text-to-speech (TTS) aims to generate natural and understandable audio based on given text content\cite{zhang2023survey}. 
Tacotron 1/2\cite{wang2017tacotron, 8461368} are early RNN-based TTS systems that significantly improved speech quality compared to previous methods.
Transformer-based TTS systems\cite{vaswani2017attention, li2019neural, lancucki2021fastpitch, lim2020jdi} excel at modeling long-dependency speech and text sequences.
Fastspeech\cite{ren2019fastspeech, ren2020fastspeech} enhanced the robustness of TTS-generated audio by reducing word skipping and repetition with an external aligner.
Glow-TTS\cite{kim2020glow} is a flow-based model that searches the most likely monotonic alignment between text and speech latent representations without needing external guidance.
Despite their satisfactory performance, current diffusion and flow matching \cite{dao2023flow, lipman2022flow} based models achieve better naturalness, speaker similarity, and sound quality.
Diff-TTS\cite{jeong2021diff} is one of the first diffusion-based TTS models, using a denoising diffusion framework to convert noisy signals into Mel-spectrograms to generate high-fidelity audio.
In addition, diffusion-based models can produce audio quality that is indistinguishable from human speech, even replicating emotions and styles to a lifelike degree\cite{shen2023naturalspeech, li2024styletts, zhang2024speechgpt, huang2022prodiff, kim2022guided, chen2020wavegrad, kong2020diffwave, lam2022bddm, lee2021priorgrad}.
Flow matching (FM) based models primarily accelerate training and inference speed. They enable accurate synthesis with fewer steps\cite{le2024voicebox, mehta2024matcha, kim2023pflow}.
However, the above-mentioned advancements in TTS technology also raise security concerns, as they provide malicious attackers with new speech synthesis tools that can lead to large-scale misuse.

Spoof detection \cite{wu2023defender, wu15e_interspeech, liao2022adversarial, chen2024singing, chen2024neural, pan24c_interspeech, 10447811} aims to distinguish genuine and spoofed utterances. 
To advance the development of anti-spoofing models, a large number of anti-spoofing challenges and datasets have been proposed so far \cite{wu15e_interspeech, delgado18_odyssey, Nautsch_2021,wu2024codecfake,10155166,frank2021wavefake,muller2022does,muller2024mlaad}.
In recent years, significant progress has been made in developing high-performance anti-spoofing models for traditional speech synthesis systems.
However, diffusion and FM based models are relatively new, and it remains uncertain whether the most advanced anti-spoofing models can effectively counter these types of synthetic speech.

In this paper, we introduce the Diffusion and Flow-matching based Audio Deepfake Dataset (DFADD), which comprehensively collects various advanced Diffusion and Flow-matching TTS models. 
The DFADD dataset comprises five diverse and mainstream open-source Diffusion and FM based TTS models.
Additionally, we conduct a comprehensive analysis, meticulously evaluating the effectiveness of cutting-edge anti-spoofing models when confronted with synthesized speech generated by these advanced Diffusion and Flow-matching TTS models. 
Moreover, we utilize the DFADD dataset to develop significantly enhanced anti-spoofing models for effectively detecting spoofed audio generated by diffusion or flow matching based TTS systems.

We observe that: (1) Models trained on the ASVspoof dataset face challenges in detecting speech clips generated by advanced diffusion and FM based TTS systems. (2) Our proposed DFADD dataset significantly improves the models' ability to handle synthesized speech from current various state-of-the-art (SOTA) diffusion and FM based TTS systems (Compared to training on ASVspoof datasets, the models trained on DFADD subsets achieve an average equal error rate (EER) reduction of over 47\%). 

We will soon release the data ~\footnote{\href{https://github.com/isjwdu/DFADD}{https://github.com/isjwdu/DFADD}} and hope this study and the DFADD dataset can reduce malicious attacks from advanced diffusion and FM based TTS systems. Our audio samples can be found on the demo page~\footnote{\href{https://github.com/DFADD-Dataset/DFADD_demo_pages/}{https://github.com/DFADD-Dataset/DFADD\_demo\_pages/}}.



\section{Related Work}

The development of anti-spoofing models requires extensive and robust datasets as training data. 
We will elaborate on existing training datasets and defense models in related work.

\subsection{Audio Anti-Spoofing Dataset}
Several audio anti-spoofing datasets have been released using various deepfake techniques, including generative models, partial spoofs, multimodal deepfakes, and multi-language spoofing audio. 
We introduce the audio deepfake datasets containing English speakers, with details shown in Table~\ref{tab:comparing_datasets}.

\textbf{ASVspoof19-LA} \cite{Nautsch_2021} contains spoofed audios generated from TTS and Voice Conversion (VC). 
All of them are from the VCTK dataset \cite{Yamagishi2019CSTRVC}. 
The ASVspoof2019-LA evaluation set contains 13 unknown TTS and VC algorithm-generated spoofed speech to verify the generalization of anti-spoofing detection algorithms.

\textbf{ASVspoof21-DF} \cite{10155166} is an audio dataset generated by more than 100 TTS and VC methods and includes different compression algorithms and source domains. 
ASVspoof-DF simulates the processing of different lossy codecs in real situations when handling the dataset.

\textbf{ASVspoof21-LA} \cite{10155166} includes the training and development sets of ASVspoof2019-LA and evaluation set of ASVspoof2021-LA. 
The evaluation set of ASVspoof2021-LA has been processed by real phone systems with various codecs, transmission channels, bit rates, and sampling rates.

\textbf{WaveFake} \cite{frank2021wavefake} has 117,985 spoofed speech clips. 
The bonafide speech clips are collected from LJspeech\footnote{\href{https://keithito.com/LJ-Speech-Dataset/}{https://keithito.com/LJ-Speech-Dataset/}} and JSUT \cite{sonobe2017jsut} dataset. Its 10 subsets are generated from 5 different GAN-based TTS models and one flow-based generative model across two languages.

\textbf{In-The-Wild (ITW)} \cite{muller2022does} contains 37.9 hours of audio recordings of celebrities and politicians, 17.2 hours of which are faked. 
There may be background noise since the recorded audio is publicly available on the Internet.

\textbf{TIMIT-TTS}\cite{salvi2023timit} is an audio dataset that uses the VidTIMIT \cite{7232917} dataset as a reference, which can be used for multimodal synthetic media detection or as an audio deepfake dataset only.
The video in VidTIMIT is split into audio content and visual content, and the deepfake audio is generated through three steps applied to the audio content.
First, the original audio is transcribed into text by a speech-to-text algorithm.
Second, spoof audio is generated from text using 12 existing TTS models.
Finally, the spoof audio is synchronized with the original audio.


\textbf{MLADD} \cite{muller2024mlaad} is a multi-language audio anti-spoofing dataset, utilizing 54 TTS models built from 21 different architectures, and generating 163.9 hours of synthetic voice across 23 different languages. 
The dataset is introduced because of the language bias present in deepfake audio datasets, most of them predominantly consist of English speech. 
By incorporating multilingual audio samples, detection models can enhance their ability to generalize across datasets, thus more advantageously combat audio spoofing and deepfakes.

However, the aforementioned datasets do not consider the newly emerged diffusion and FM based TTS models.

\subsection{Anti-spoofing}

The anti-spoofing model is designed to differentiate between genuine and spoofed utterances, mitigating the impact of synthetic speech.
AASIST \cite{jung2022aasist} is one of the SOTA anti-spoofing models. 
It takes Rawnet2 \cite{tak2021end} as its speech encoder to extract features and employs two graph modules for spectral and temporal domain modeling.
In addition, it utilizes max graph operations with heterogeneous graph modeling via HS-GAL \cite{wang2019heterogeneous} layers, and achieves final classification through an output layer after element-wise maximum and node value aggregation.
AASIST-L uses several model compression techniques to reduce its size by 70\% compared to the AASIST model while keeping the overall architecture unchanged, alleviating the overfitting issues.
We use AASIST-L as our backbone trained on different subsets of DFADD. 
This is because AASIST-L is less prone to overfitting on ASVspoof2019 dataset.
More detailed experimental setups will be explained in \textbf{Section~\ref{subsec: anti-spoofing-setup}}.

\captionsetup[table]{skip=-8pt}
\begin{table}[t]
\caption{Comparison of DFADD with other deepfake datasets containing English speakers \cite{li2024audio}. None means no detailed information is provided.}
\vspace{1em}
\centering
\fontsize{10}{10}\selectfont
\setlength\tabcolsep{2pt}
\resizebox{0.48\textwidth}{!}{
\begin{tabular}{ccccccc}
  \toprule
  Dataset & Type & Methods & Speakers & Language & Bonafide & Spoofed \\
  \midrule
  \midrule
  \makecell{ASVspoof19 \\ LA} & \makecell{TTS \\ VC} & 19 & 48 & English & 10,256 & 90,192 \\
  \midrule
  \makecell{ASVspoof21 \\ DF} & \makecell{TTS \\ VC} & 100+ & 93 & English & 14,869 & 519,059 \\
  \midrule
  \makecell{ASVspoof21 \\ LA} & \makecell{TTS \\ VC} & 19 & 67 & English & 14,816 & 133,360 \\
  \midrule
  \makecell{WaveFake} & TTS & 7 & 2 & \makecell{English \\ Japanese} & 0 & 117,985 \\
  \midrule
  \makecell{TIMIT} & TTS & 12 & 46 & English & 0 & 5,160 \\
  \midrule
  \makecell{ITW} & None & None & 58 & English & 19,963 & 11,816 \\
  \midrule
  \makecell{MLAAD} & TTS & 54 & None & \makecell{English, \\ other 22} & 0 & 76,000 \\
  \midrule
  \midrule
  \makecell{\textbf{DFADD}} & TTS & 5 & 109 & English & 44,455 & 163,500 \\
  \toprule
\end{tabular}
}
\label{tab:comparing_datasets}
\end{table}

\section{DFADD dataset}

\begin{figure}[t]
\centering
\includegraphics[width=8.5cm]{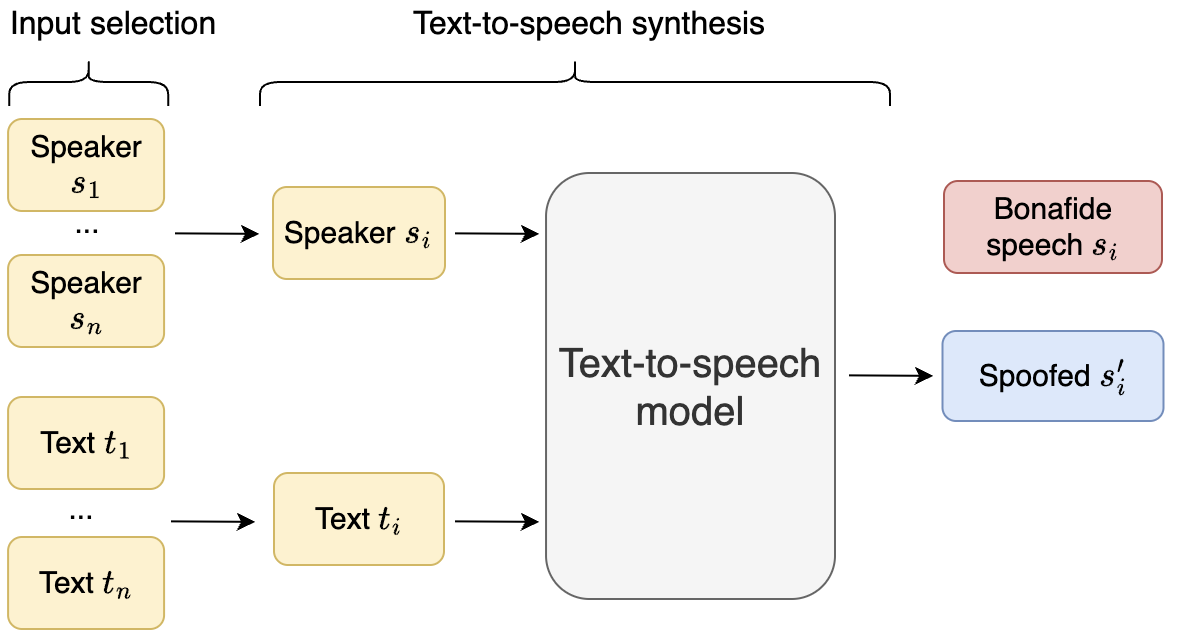}
\vspace{-1mm}
\caption{The pipeline of data generation for DFADD.} 
\label{fig:flow}
\end{figure}

This section describes the creation of the DFADD dataset, its design principles, and the reasoning behind its development.
The pipeline for generating our dataset is shown in Fig.\ref{fig:flow}. 
It consists of two stages: input selection and text-to-speech synthesis.
In the input selection stage, we obtain the target speaker prompts $s$ and text prompts $t$.
In the text-to-speech synthesis stage, each selected speaker prompt $s_i$ and selected text prompt $t_i$ will be input to the TTS model to generate spoofed audio.
We used 5 different diffusion and FM-based TTS models, which will be introduced in Section.\ref{TTSmodels}
There is a one-to-one correspondence between bonafide and spoofed speakers. 
In other words, both bonafide and spoofed speakers have the same speaker identity, with the difference being that one is authentic while the other is synthesized through TTS.
\vspace{-1.5mm}

\subsection{Input selection}
\vspace{-1.5mm}
\textbf{Text selection.} To prevent text data leakage from models pre-trained on VCTK, we avoided using the same text prompts from VCTK, and instead used LJspeech to get our text prompts.
To ensure the audio duration and quality are similar to VCTK samples, we removed sentences with complex words (such as names and special nouns) and selected sentences with 5 to 10 words.
Ultimately, we chose 300 sentences that met these criteria to serve as text prompt inputs for the subsequent stage of speech synthesis in TTS systems.

\textbf{Speaker selection.} We use the VCTK dataset, which includes 109 speakers, as our input.
Each speaker undergoes inference on the 300 sentences selected during \textbf{Text selection} to generate the spoofed audio in the synthesis stage.

\subsection{Text-to-speech model}
\label{TTSmodels}
We leveraged the released checkpoints trained on VCTK or zero-shot models using diffusion and FM based methods.
We selected 5 different TTS systems as the backbone for our dataset generation.
Diffusion-based TTS systems include Grad-TTS \cite{popov2021grad}, NaturalSpeech 2 \cite{shen2023naturalspeech}, and Style-TTS 2 \cite{li2024styletts}.
FM-based TTS systems include Matcha-TTS \cite{mehta2024matcha} and PFlow-TTS \cite{kim2023pflow}. 
To simplify the description, let ``D$\sim$" stands for Diffusion, and ``F$\sim$" represents Flow-matching. 

\subsubsection{Diffusion-based Text-to-speech}
\label{subsubsec: diffusionTTS}


Diffusion-based TTS models introduce noise into audio features and progressively denoise them to produce high-quality speech features or waveforms.
Their superior performance is due to their ability to model complex data distributions with fine-grained control, reducing artifacts, controlling speech emotions, and stylizing speech from text, resulting in highly natural audio. 

\textbf{D1. Grad-TTS \cite{popov2021grad}.} It encodes the text into features and aligns with the text input using the Monotonic Alignment Search algorithm, creating a monotonic mapping between the text and the mel-spectrogram.
The diffusion process generates mel-spectrograms from Gaussian noise, guided by a noise scheduling function and reversed through time-based inference to reconstruct the target distribution gradually. 


\textbf{D2. NaturalSpeech 2 \cite{shen2023naturalspeech}.} In the training process, NaturalSpeech 2 converts the input speech waveform into quantized latent vectors. Then, a diffusion model predicts these latent vectors from the text input. The model conditions on the output of phoneme encoder, duration predictor, and pitch predictor. During inference, the diffusion model first generates the latent vectors from the text or phoneme sequence and then converts these latent vectors into the final speech waveform using the decoder in the neural audio codec. 

\textbf{D3. Style-TTS 2 \cite{li2024styletts}.} It uses a text encoder to convert the input text into phoneme representations. The diffusion model samples a style vector from a latent random variable conditioned on the input text, modeling diverse speech styles. The style vector is fed into the speech decoder, which combines it with the phoneme representations, pitch curve, and energy curve to generate the final speech waveform directly. To achieve efficient generation, StyleTTS 2 uses large pre-trained speech language models (e.g., WavLM) as discriminators and introduces differentiable duration modeling to enhance speech naturalness and generation quality.



\subsubsection{Flow-matching based Text-to-speech}
\label{subsubsec: flow-matchingTTS}

FM-based TTS models further enhance the efficiency.
It eliminates the need for numerically solving the reverse-time stochastic differential equation, which requires many steps.
After obtaining the acoustic features, these models aim to directly model the vector field implied by an arbitrary ordinary differential equation (ODE). 
All FM-based models consider linearized sampling trajectories and minimize transmission costs from data distribution or noise, thereby finding a more straightforward path from source to target, resulting in higher-quality synthesis with fewer steps.

\textbf{F1. Matcha-TTS \cite{mehta2024matcha}.} Matcha-TTS employs a text encoder to convert the input text into a sequence of phonemes, capturing textual information.
A duration predictor estimates the duration for each phoneme to ensure the synthesized speech is synchronized using the input text. 
It then employs a conditional flow matching approach to train the whole model, optimizing the path from latent space to the data distribution, thereby reducing the number of steps needed for synthesis. 
Finally, the duration predictor’s output with the diffusion process generates mel-spectrograms from noise and uses a neural decoder to convert these mel-spectrograms into the final speech waveforms.

\textbf{F2. PFlow-TTS \cite{kim2023pflow}.} PFlow-TTS is a zero-shot TTS model that generates high-quality speech for unseen speakers using minimal training data.
It consists of a speech-prompted text encoder that combines a short speech prompt with text input to produce a speaker-conditioned text representation.
This representation is used by the flow-matching generative decoder to synthesize speech, converting the text to a mel-spectrogram and then to a waveform.
PFlow-TTS achieves superior speed and data efficiency by avoiding autoregressive components and neural codecs, using flow matching for faster and more direct speech synthesis.
This method provides significant improvements in inference speed and speaker adaptation, maintaining high speech quality with reduced data and simpler training.

\subsubsection{Text-to-speech synthesis}

For D2, D3, and F1, we used models pre-trained on VCTK to perform inference on different speakers.
We trained D1 and F2 from scratch to adapt to VCTK speakers.
F2 is an unofficial implementation.
F1 is the only officially open-sourced TTS system that uses the FM method and supports inference for VCTK speakers.

\textbf{D1. Grad-TTS audio synthesis.} During the training phase, we followed the default Grad-TTS hyperparameter settings.
We trained Grad-TTS for 1000 epochs on a V100-32G GPU, with a batch size of 16 and a sample rate of 22,050 Hz. 
During the inference phase, we replaced the vocoder provided by Grad-TTS with HiFi-GAN, which is pre-trained on VCTK. 
Additionally, we changed the diffusion time steps and temperature to 70 and 3, respectively. 

\textbf{D2. NaturalSpeech 2 audio synthesis.} We used a zero-shot approach with a prompt speech and the text mentioned in \textbf{Input selection} to generate a spoofed speech.
The prompt speech we used is utterance number 016 of bonafide audio from each speaker.
We use the unofficial checkpoint \footnote{\href{https://github.com/CODEJIN/NaturalSpeech2}{https://github.com/CODEJIN/NaturalSpeech2}} pre-trained on the VCTK for 306K steps with V100-32G GPU.

\textbf{D3. Style-TTS 2 audio synthesis.} StyleTTS2 also uses a zero-shot approach with a prompt speech and text to generate spoofed D3 subset, similar to NaturalSpeech 2.
We use the checkpoint pre-trained on the LibriTTS dataset and set the parameters $\alpha$ and $\beta$ in Style-TTS 2 both to 0, making the generated spoof speech as similar as possible to the original.

\textbf{F1. Matcha-TTS audio synthesis.} We used the official checkpoint pre-trained on VCTK to generate the F1 subset. For inference, we used a V100 GPU, with the temperature set to 0.667 and the ODE step set to 10.

\textbf{F2. PFlow-TTS audio synthesis.}  During the training phase, we follow PFlow-TTS’s default hyperparameter settings. We trained for 1100 epochs on a GPU V100 32G, and the batch size was 16.
During the inference phase, we replace the vocoder provided by the unofficial PFlow-TTS\footnote{\href{https://github.com/p0p4k/pflowtts\_pytorch}{https://github.com/p0p4k/pflowtts\_pytorch}} with HiFi-GAN \cite{kong2020hifi}, which is pre-trained on VCTK. 
Since some poor-quality real audio is removed from the VCTK, we use the bonafide audio with utterance number 013 as the prompt speech of speakers p292 and p318. 
For other speakers, the utterance number of prompt speech is 003. 
Our ODE steps and temperature are the same as F1's settings.

\subsection{Dataset comparison}
\label{dataset_comparison}

We divided 109 speakers into three speaker-disjoint sets for training, validation, and testing. Speakers for validation data are p226 and p229, while speakers p227 and p228 are used for testing. The remaining speakers are allocated for training. The sample rate of all audio files is set to 16,000 Hz. 

We generate 163,500 TTS-based spoofed speech clips totaling 179.88 hours from the bonafide speech clips, with an average length of 4.01 seconds. Table~\ref{tab:dfadd_dataset} shows a detailed summary of subsets. In comparison with existing mainstream audio anti-spoofing datasets, which primarily use TTS and VC methods, our dataset focuses solely on TTS systems. While previous TTS anti-spoofing datasets were generated using traditional neural network methods (e.g., Flow-based, GAN-based), these methods are inferior to diffusion-based and FM-based approaches in generation quality. Furthermore, DFADD features the largest number of speakers among anti-spoofing datasets in Table~\ref{tab:comparing_datasets}, and the speech clips we generated far exceed the number of spoofed speech clips in other TTS-only datasets.



\section{experimental setup}

\subsection{Anti-spoofing model setup}
\label{subsec: anti-spoofing-setup}
We use one of SOTA deepfake detection models, AASIST-L \cite{jung2022aasist}~\footnote{\href{https://github.com/clovaai/aasist}{https://github.com/clovaai/aasist}}, as our backbone for training anti-spoofing models.
We chose AASIST-L because it is less prone to overfitting on the ASVspoof dataset.

\captionsetup[table]{skip=0pt}
\begin{table}[t]
\caption{Comparison between DFADD different generation pipelines. D1 refers to the GradTTS. D2 signifies NaturalSpeech 2. D3 represents the StyleTTS 2. F1 means the Matcha-TTS. F2 represents the PFlow-TTS. Train, valid, and test represent the average duration (seconds), respectively.}
\vspace{0.5em}
\centering
\fontsize{10}{10}\selectfont
\setlength\tabcolsep{9pt}
\resizebox{0.48\textwidth}{!}{%
\begin{tabular}{lcccccc}
  \toprule
  Subsets & Methods  & Source & train & valid & test \\
  \midrule
  \midrule
  \multicolumn{1}{c}{\centering D1} & Diffusion  & Train & 3.06 & 3.06 & 3.06 \\
  \midrule
  \multicolumn{1}{c}{\centering D2} & Diffusion  & Pretrain & 5.63 & 5.67 & 5.84 \\
  \midrule
  \multicolumn{1}{c}{\centering D3} & Diffusion  & Pretrain & 3.84 & 3.97 & 4.10 \\
  \midrule
  \multicolumn{1}{c}{\centering F1} & Flow matching  & Pretrain & 2.98 & 2.88 & 3.10 \\
  \midrule
  \multicolumn{1}{c}{\centering F2} & Flow matching  & Train & 4.27 & 4.24 & 4.42 \\
  \midrule
  \midrule
  \multicolumn{1}{c}{\centering DFADD} & -  & - & 3.83 & 3.85 & 4.01 \\  
  \toprule
\end{tabular}%
}
\label{tab:dfadd_dataset}
\end{table}
\vspace{1em}
For the ASVspoof dataset, we utilize the author’s released checkpoints after their thorough hyperparameter search.
For DFADD, our model used the Adam optimizer with a learning rate of 0.001 and a batch size of 24, trained on a V100 32G GPU.
During training, one of the DFADD subsets is used as spoofed audio, and the corresponding bonafide VCTK utterances are combined as training data.
We use the corresponding DFADD validation subset for model selection.

\subsection{Evaluation setup}


We consider two evaluation scenarios: the seen scenario and the unseen scenario.
(1) The seen scenario involves evaluating the anti-spoofing model on the evaluation set of each DFADD subset.
This means the model has been exposed to the same distribution of datasets during training and has learned the features of the corresponding subset. 
(2) The unseen scenario involves evaluating the models on audio samples collected from demo pages of various TTS systems. 
Since the models were trained on the DFADD subsets, they did not learn the features from these collected audio samples.
These TTS systems in unseen scenario include VoiceBox \cite{le2024voicebox}, VoiceFlow \cite{guo2024voiceflow} NaturalSpeech 3 \cite{ju2024naturalspeech}, CMTTS \cite{li-etal-2024-cm}, DiffProsody \cite{oh2024diffprosody}, and DiffAR \cite{benita2023diffar}.
VoiceFlow and VoiceBox use FM-based methods, while the others use diffusion-based methods.
The ground truth data comes from actual recordings or speaker prompts, and the model-generated audio is classified as spoofed data.
By evaluating on these unseen datasets, we assess the generalization performance of our models.

\section{experiments results}
\subsection{Audio quality assessment}
\label{subsec: audio_assessment}
\begin{figure}[t]
\centering
\includegraphics[width=8cm]{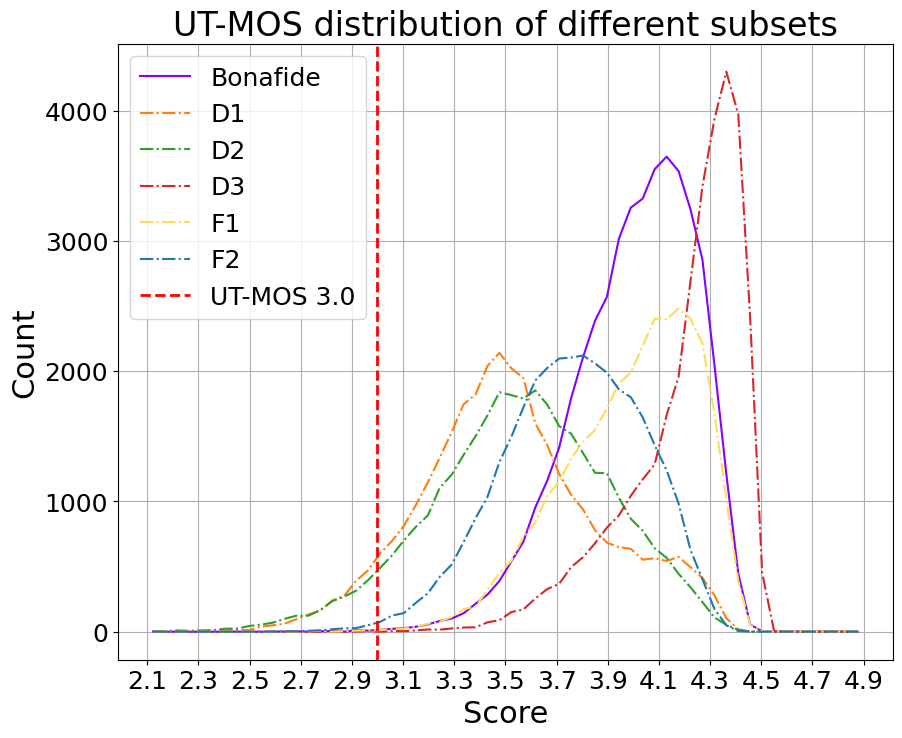}
\vspace{-3mm}
\caption{MOS distribution of spoofed audio generated by different TTS models (higher means more natural).} 
\label{fig:mos_results}
\end{figure}


We leverage the UT-MOS \cite{saeki2022utmos} to assess the quality of our synthesized audio in the DFADD dataset. 
MOS usually ranges from 1 to 5, where higher scores represent better natural synthesis quality.
The detailed MOS distribution of DFADD is shown in Fig-\ref{fig:mos_results}. 
Over 97\% of DFADD speech clips (including bonafide and spoofed) have an MOS of 3.0 or above. 
The quality of the spoofed audio generated by D3 is especially natural, with most MOS greater than 4.0, making it comparable to bonafide audio.
While UT-MOS has known limitations and inherent biases \cite{zielinski2008some, rosenberg2017bias, cooper2023investigating}, the results still indicate that most quality of spoofed audio samples in DFADD are close to genuine audio, highlighting the potential misuse of diffusion and FM-based TTS models in malicious attacks.

\subsection{Seen scenario cross-testing evaluation}
\label{subsec: seen_scenario}
\definecolor{Gray}{gray}{0.85}
\newcolumntype{g}{>{\columncolor{Gray}}c}
\captionsetup[table]{skip=10pt}
\begin{table*}[t]

\fontsize{9}{10}\selectfont
\centering
\caption{Performance comparison of spoofed speech detection (EER) between models trained on ASVspoof and DFADD. Models surpassing those trained on ASVspoof are emphasized in bold. The top-performing models feature a gray background.}
\setlength\tabcolsep{10pt}
\resizebox{\textwidth}{!}{%
\begin{tabular}{cccccccc}
  \toprule
  & \multicolumn{2}{c}{ASVspoof (All)} & \multicolumn{5}{c}{DFADD} \\
  \cmidrule(lr){2-3} \cmidrule(lr){4-8}
  & AASIST & AASIST-L & D1 & D2 & D3 & F1 & F2 \\
  \midrule
  \midrule
  VoiceBox \cite{le2024voicebox} & 42.59 & 47.62 & 53.77 & 64.70 & 50.13 & \cellcolor{Gray}\textbf{36.69} & 60.80 \\
  \midrule
  VoiceFlow \cite{guo2024voiceflow} & 50.41 & 41.33 & \textbf{34.70} & \textbf{33.06} & \textbf{33.06} & 42.96 & \cellcolor{Gray}\textbf{24.80} \\
  \midrule
  NaturalSpeech3 \cite{ju2024naturalspeech} & 24.50 & 25.50 & 31.63 & 59.69 & 62.25 & \cellcolor{Gray}\textbf{18.38} & \textbf{24.50} \\
  \midrule
  CMTTS \cite{li-etal-2024-cm} & 56.54 & 43.46 & \textbf{20.26} & \textbf{10.13} & \textbf{10.13} & \cellcolor{Gray}\textbf{0.00} & \cellcolor{Gray}\textbf{0.00} \\
  \midrule
  DiffProsody \cite{oh2024diffprosody} & 37.50 & 35.94 & 62.50 & \textbf{28.13} & \cellcolor{Gray}\textbf{25.00} & \cellcolor{Gray}\textbf{25.00} & \cellcolor{Gray}\textbf{25.00} \\
  \midrule
  DiffAR \cite{benita2023diffar} & 53.72 & 69.42 & 74.73 & \textbf{50.53} & \textbf{27.39} & \cellcolor{Gray}\textbf{25.27} & \textbf{4.26} \\
  \midrule
  \midrule
  Average & 44.21 & 43.88 & 46.26 (\textcolor{blue}{+2.38}) & 41.04 (\textcolor{red}{-2.84}) & 34.66 (\textcolor{red}{-9.22}) & 24.72(\textcolor{red}{-19.16}) & 23.22 (\textcolor{red}{-20.66}) \\
  \bottomrule
\end{tabular}%
}
\vspace{0.5em}
\label{tab:half_unseen}
\end{table*}

\begin{figure}[t]
\centering
\includegraphics[width=8.5cm]{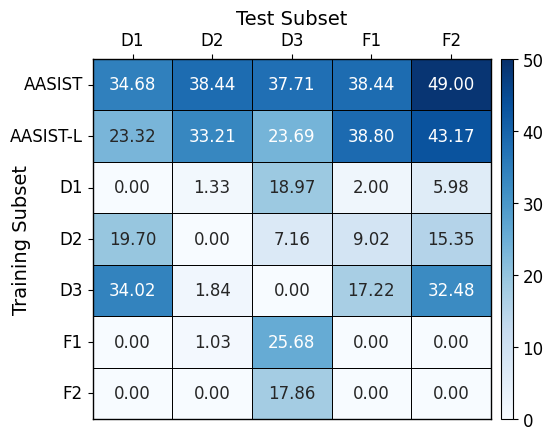}
\vspace{-3mm}
\caption{Cross-testing EER results of anti-spoofing models on DFADD test subsets. The evaluation metric is equal error rate (EER), where lower is better.} 
\label{fig:seen_results}
\end{figure}


The Fig.~\ref{fig:seen_results} shows the cross-testing results followed by the dataset splitting method in Section~\ref{dataset_comparison}. 
The rows in the figure represent subsets of training data from DFADD and ASVspoof, while the columns correspond to subsets of testing data from DFADD. Our observations are as follows:


(1) Most testing subsets show significantly high EERs, typically above 30\%, for the AASIST and AASIST-L models trained on the ASVspoof dataset. This indicates that these models struggle to distinguish speech generated by diffusion-based and FM-based TTS systems.

(2) From the perspective of the horizontal axis, models training on subsets from the FM-based TTS pipeline (F1, F2) perform very well on the diffusion-based test subsets except for D3, with EERs very close to 0. 
This indicates that the anti-spoofing model trained on the FM-based audio deepfake dataset has better generalization performance compared to the model trained on the diffusion-based audio deepfake dataset.

(3) From the perspective of the vertical axis, training on a particular DFADD subset consistently results in the lowest EER for its corresponding testing subset.
For instance, training on subset D1 yields the lowest EER when testing on D1, and similarly for subset F1.
Additionally, D3 is the most difficult subset to fit in each training and testing scenario. This may be because the audio of D3 is so realistic that the models trained on other subsets cannot distinguish it. This is also indirectly indicated by the higher MOS scores of D3 compared to the other subsets in Fig.\ref{fig:mos_results}.

\subsection{Unseen scenario cross-testing evaluation}
\label{subsec: unseen_scenario}

Table~\ref{tab:half_unseen} presents the performance of AASIST-L models trained on various subsets and evaluated on the unseen scenario.
The columns show the training sets used to develop each model. 
For the ASVspoof dataset, all available data were used to train two model variants: AASIST and AASIST-L.
For the DFADD, each column indicates the training subset used for the AASIST-L model. 
The rows indicate the sources of the testing sets. The following observations were made:

(1) The anti-spoofing models trained with the ASVspoof dataset exhibit notably poor performance on the unseen evaluation dataset.
This pronounced discrepancy likely arises because the ASVspoof dataset mainly contains speech clips generated by traditional TTS and VC methods, which differ significantly from the diffusion and FM based methods in DFADD. 
This difference highlights the urgent need for datasets generated by these advanced methods to improve the robustness of anti-spoofing models.

(2) From the perspective of each unseen evaluation set, the EER of the model trained on a single subset is generally lower than when trained on ASVSpoof in most unseen evaluation datasets.
Notably, the CMTTS models show a significant decrease in EER regardless of the subset used for training.
Additionally, models trained on FM-based TTS subsets exhibit the highest degree of generalizability, significantly reducing their EER in most unseen scenarios.

(3) From the perspective of average EERs on individual DFADD subsets, anti-spoofing models trained on DFADD subsets show a high effectiveness in detecting spoofing in unseen and similar methods (diffusion and FM based TTS) datasets.
Specifically, the average EERs of models trained on F1 and F2 are reduced by 19.16 and 20.66, respectively, compared to the baseline AASIST-L (trained on ASVspoof).
In addition, the reduction achieved by anti-spoofing models trained on FM-based audio samples is significantly greater than that achieved by models trained on diffusion-based subsets.
These findings indicate that models trained on FM-based subsets exhibit better generalization capabilities.

\section{conclusion}

In this study, we assembled DFADD, the first dataset that includes spoofed speech generated specifically using advanced diffusion and FM based TTS models.
We verified that the spoofed audio generated by these models has a highly natural quality.
Our extensive experiments demonstrate that anti-spoofing models trained on the ASVspoof dataset struggle to detect spoofs from diffusion and FM based TTS models, but the DFADD dataset significantly enhances their performance.
The average EER of an anti-spoofing model on unseen scenarios was reduced by more than 47\% due to train on DFADD subsets.
All codes and data will soon be released to help resist malicious attacks from advanced diffusion and FM based speech synthesis systems.


\section{Acknowledgements}
This work was partially supported by the National Science and Technology Council, Taiwan (Grant no. NSTC 112-2634-F-002-005, Advanced Technologies for Designing Trustable AI Services). We also thank the National Center for High-performance Computing (NCHC) of National Applied Research Laboratories (NARLabs) in Taiwan for providing computational and storage resources.

\bibliographystyle{IEEEbib}
\bibliography{Template_Regular}

\end{document}